\documentclass[twocolumn,showpacs,preprintnumbers,amsmath,amssymb]{revtex4-1}
\usepackage{amsfonts}
\usepackage{amsmath}
\usepackage{graphicx}
\usepackage{dcolumn}
\usepackage{bm}
\usepackage{overpic}
\usepackage{booktabs}
\usepackage{color}

\begin{document}
\title{Enhancing network robustness for malicious attacks}
\author{An Zeng}\email{an.zeng@unifr.ch}
\author{Weiping Liu}
\affiliation{Department of Physics, University of Fribourg, Chemin du Mus\'{e}e 3, CH-1700 Fribourg, Switzerland}
\date{\today}

\begin{abstract}
In a recent work [Proc. Natl. Acad. Sci. USA 108, 3838 (2011)], the authors proposed a simple measure
for network robustness under malicious attacks on nodes. With a greedy algorithm, they found the optimal structure with respect to
this quantity is an onion structure in which high-degree nodes form a core surrounded by rings of nodes with decreasing degree. However, in real networks the failure can also occur in links such as dysfunctional power cables and blocked airlines. Accordingly, complementary to the node-robustness measurement ($R_{n}$), we propose a link-robustness index ($R_{l}$). We show that solely enhancing $R_{n}$ cannot guarantee the improvement of $R_{l}$. Moreover, the structure of $R_{l}$-optimized network is found to be entirely different from that of onion network. In order to design robust networks resistant to more realistic attack condition, we propose a hybrid greedy algorithm which takes both the $R_{n}$ and $R_{l}$ into account. We validate the robustness of our generated networks against malicious attacks mixed with both nodes and links failure. Finally, some economical constraints for swapping the links in real networks are considered and significant improvement in both aspects of robustness are still achieved.

\end{abstract}
\maketitle

\section{Introduction}
The security of the infrastructure in modern society is of great importance. Systems like internet, power grids, transportation and fuel distribution networks need to be robust and capable of surviving from random failures or intentional attacks~\cite{Nature406378}. Many processes taking place on networks might be significantly influenced and showing low degree of tolerance to damage on their structures~\cite{PNAS10622073,PRE85036101}. Examples of such processes in nature and society include epidemic spreading~\cite{PRL863200,NP6888}, synchronization~\cite{PR46993,PRE83045101R,EPL8748002}, random walks~\cite{PRL92118701,PNAS1051118}, traffic~\cite{PRL104018701,EPL8958002} and opinion formation~\cite{PRE72046113,RMP81591}. Therefore, the robustness for different network structures was intensively studied in the past decade~\cite{PRL855468,PRL854626,EPJB52563,PRL90068701,PRL102018701}. It is also revealed that the shortest path~\cite{PRL87198701} and graph spectrum~\cite{CMJ23298} can be employed to estimate the network robustness. Moreover, interdependent network~\cite{Nature4641025,PRE83065101R} is proposed to model the catastrophic cascade of failures in real systems.

In a recent work, a new measure for network robustness under malicious attack on nodes is proposed~\cite{PNAS1083838}. This measurement, which we call node-robustness in this paper, considers the size of the largest component during all possible malicious attacks, namely $R_{n}=\frac{1}{N}\sum^{1}_{q=1/N}S(q)$, where $N$ is the number of nodes in the network and $S(q)$ is the fraction of nodes in the largest connected cluster after removing $qN$ nodes. The normalization factor $1/N$ makes robustness of networks with different sizes comparable. A robust network is generally corresponding to a large $R_{n}$ value. With this measurement, a greedy algorithm is designed to enhance the node-robustness in real systems and large improvement is observed even a small number of links are modified. Moreover, the optimal structure for node-robustness is found to be an onion structure in which high-degree nodes are highly connected with rings of nodes with decreasing degree surrounding. Lately, some simple methods were also proposed to generate such robust onion networks~\cite{PRE84026106}.

However, the analysis in ref.~\cite{PNAS1083838} is only based on the targeted attacks on nodes. In reality, failures can happen in connections between nodes as well~\cite{EPJB52563}. For example, the power cables can be dysfunctional and some airlines can be blocked due to the terrible weather or terrorist attacks. In this paper, we propose a link-robustness index ($R_{l}$) to measure the ability of network to resist link failures. We find that solely enhancing $R_{n}$ cannot always improve $R_{l}$ and the network structure for optimal $R_{l}$ is far different from the onion network. In order to design robust network resistant to different kinds of malicious attacks, we propose a greedy algorithm aiming for both $R_{n}$ and $R_{l}$ improvement. To validate the robustness of the resultant networks, we examined them against more realistic attack strategy which combines both nodes and links failure. Since the manipulation of real network always confront certain economical constraints, we finally took these requirements into consideration in our method and some significant improvement in both $R_{l}$ and $R_{n}$ are still obtained.

\section{Link-robustness of networks}

Since a robust network should be able to resist the most destructive attack, we begin our analysis by comparing the harmfulness caused by different malicious attack strategies on links. The most destructive attack is supposed to destroy the most ``important" links in the networks. Like ref.~\cite{PNAS1083838}, we monitor the size of giant component to estimate how the network gets destroyed after these ``important" links are removed step by step. There are many methods to measure the ``importance" of links, here we mainly consider three indexes to identify the most important link to delete. The indexes include edge-betweenness, link clustering coefficient and degree product. We also use the random link removal as a benchmark for comparison. In order to simulate a more harmful strategy, we apply a dynamical approach in which the ``importance" of the links (i.e. edge-betweenness, link clustering coefficient and degree product) are recalculated during the attack. Fig. 1 reports how the relative size of the giant component $S(p)$ changes with the fraction of links $p$ removed by different strategies in a Barabasi-Albert (BA) network model. Obviously, the most destructive strategy is the one based on the edge-betweenness. This is because the links with high betweeness are with many shortest paths passing through. If they are cut, many nodes cannot communicate with each other and the networks are likely to break into pieces. Specifically, though the degree-based node attack strategy can make a severe damage to the network, cutting the links connecting high degree nodes leads to even less harmful effect than the random removal method to the network connectivity. This is reasonable because the hubs can be strongly connected with each other, and this is well known as the rich-club phenomenon~\cite{NaturePhys2110}.

\begin{figure}
  \center
  \includegraphics[width=7cm]{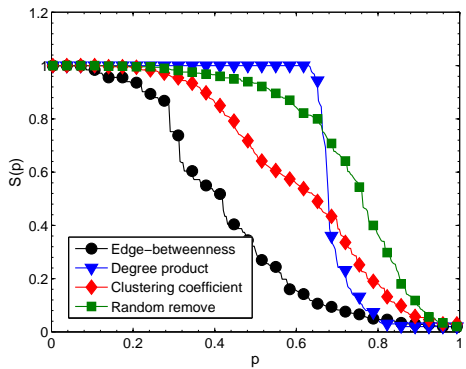}
\caption{(Color online) The change of the relative size of the giant component $S(p)$ with the fraction of links $p$ removed by different strategies in Barabasi-Albert (BA) networks. The original BA network is with $N=100$ and $\bar{k}=6$. The results are averaged over $100$ independent realizations.}
\label{fig1}
\end{figure}

\begin{figure}
  \center
  \includegraphics[width=7cm]{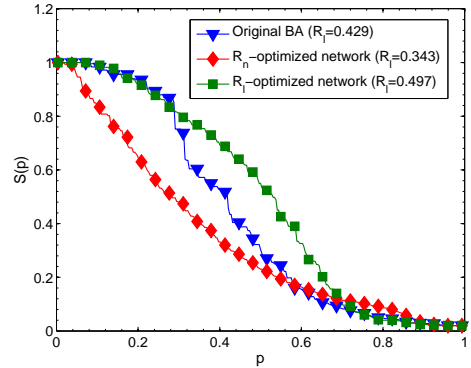}
\caption{(Color online) The $R_{l}$ of BA networks with $N=100$ and $\bar{k}=6$, the corresponding $R_{n}$-optimized and $R_{l}$-optimized networks. The results are averaged over $100$ independent realizations.}
\label{fig2}
\end{figure}

According to the analysis above, we propose a link-robustness index ($R_{l}$) based on the highest edge-betweenness attack strategy as
\begin{equation}
R_{l}=\frac{1}{E}\sum^{1}_{p=1/E}S(p),
\end{equation}
where $E$ is the total number of links. This measure captures the network response to any fraction of link removal. Apparently, if a network is robust against link attack, its $R_{l}$ should be relatively large.

In ref.~\cite{PNAS1083838}, it is found that the most robust structure for node attack is the onion-like network, which is corresponding to the topology with maximum $R_{n}$. However, it is still unclear whether this structure is tolerant to the link attack as well. In principle, a robust network should have both large $R_{n}$ and $R_{l}$ since both the nodes and links can fail due to some unexpected accidents. We therefore report the $R_{l}$ in BA networks and the corresponding onion networks in fig.2. Interestingly, despite the onion networks is resistant to malicious node attack, it is weaker than the original BA networks with respect to the intentional link attack. More specifically, the $R_{l}$ in onion network is $19.9\%$ lower than the BA model (For detail value, see table I).

Therefore, it is necessary to design a structural manipulating method to enhance the link-robustness for networks. Since changing the degree of a node is commonly assumed to be particular more expensive than changing the connections, we keep invariant the degree of each node in our algorithm. Starting from an original network, we swap the connections of two randomly chosen edges, i.e., we randomly select two edges $ab$ and $cd$ (which connect node $a$ with node $b$, and node $c$ with node $d$, respectively), then change them to $ad$ and $bc$ only if $R^{\rm new}_{l}>R^{\rm old}_{l}$. We then repeat this procedure with another randomly chosen pair of edges until no
further substantial improvement is achieved for a given large number of consecutive swapping trials (Here, we set it as $10^4$). In fig. 2, we can clearly see that the $R_{l}$ can be significantly improved by the algorithm. Compared to the original BA network, $R_{l}$ can be increased by $15.8\%$ (See table I for detail value).

\section{Improving robustness in real networks}

\begin{table*}
\caption{Properties in the different networks: Node-robustness index ($R_{n}$), Link-robustness index ($R_{l}$), the spectrum of the adjacency matrix ($\lambda_1/\lambda_2$), degree assortativity ($r$), average shortest path length ($\langle d \rangle$) and clustering coefficient ($\langle C \rangle$).}
\label{tab1}
\begin{center}
\begin{tabular}{p{2cm} p{3cm} p{1.6cm} p{1.7cm} p{1.7cm} p{1.7cm} p{1.7cm} p{1.2cm}}
\hline
\hline
Network &Algorithm &$R_{n}$ &$R_{l}$ &$\lambda_1/\lambda_2$ &$r$ &$\langle d \rangle$ &$\langle C \rangle$\\
\hline
&Original     &0.201 &0.429 &1.856 &-0.181 &2.576 &0.142\\
BA&$R_{n}$-optimized     &0.352 &0.343 &2.579 &0.158 &2.828 &0.117\\
&$R_{l}$-optimized     &0.200 &0.497 &1.891 &-0.162 &2.584 &0.137\\
&Hybrid-optimized        &0.219 &0.491 &1.898 &-0.153 &2.583 &0.133\\
~\\
&Original     &0.110 &0.244 &2.382 &-0.208 &2.738 &0.625\\
USAir&$R_{n}$-optimized     &0.293 &0.245 &5.054 &-0.148 &2.875 &0.280\\
&$R_{l}$-optimized     &0.111 &0.319 &2.631 &-0.315 &2.492 &0.480\\
&Hybrid-optimized        &0.196 &0.298 &3.018 &-0.237 &2.593 &0.429\\
~\\
&Original     &0.111 &0.093 &1.122 &0.001 &6.588 &0.123\\
Grid&$R_{n}$-optimized     &0.240 &0.173 &1.404 &0.356 &6.128 &0.015\\
&$R_{l}$-optimized     &0.125 &0.248 &1.192 &0.019 &4.974 &0.024\\
&Hybrid-optimized        &0.161 &0.237 &1.272 &0.110 &5.017 &0.031\\
\hline
\hline
\end{tabular}
\vspace*{0.0cm}
\end{center}
\end{table*}

\begin{figure}
  \center
  \includegraphics[width=7cm]{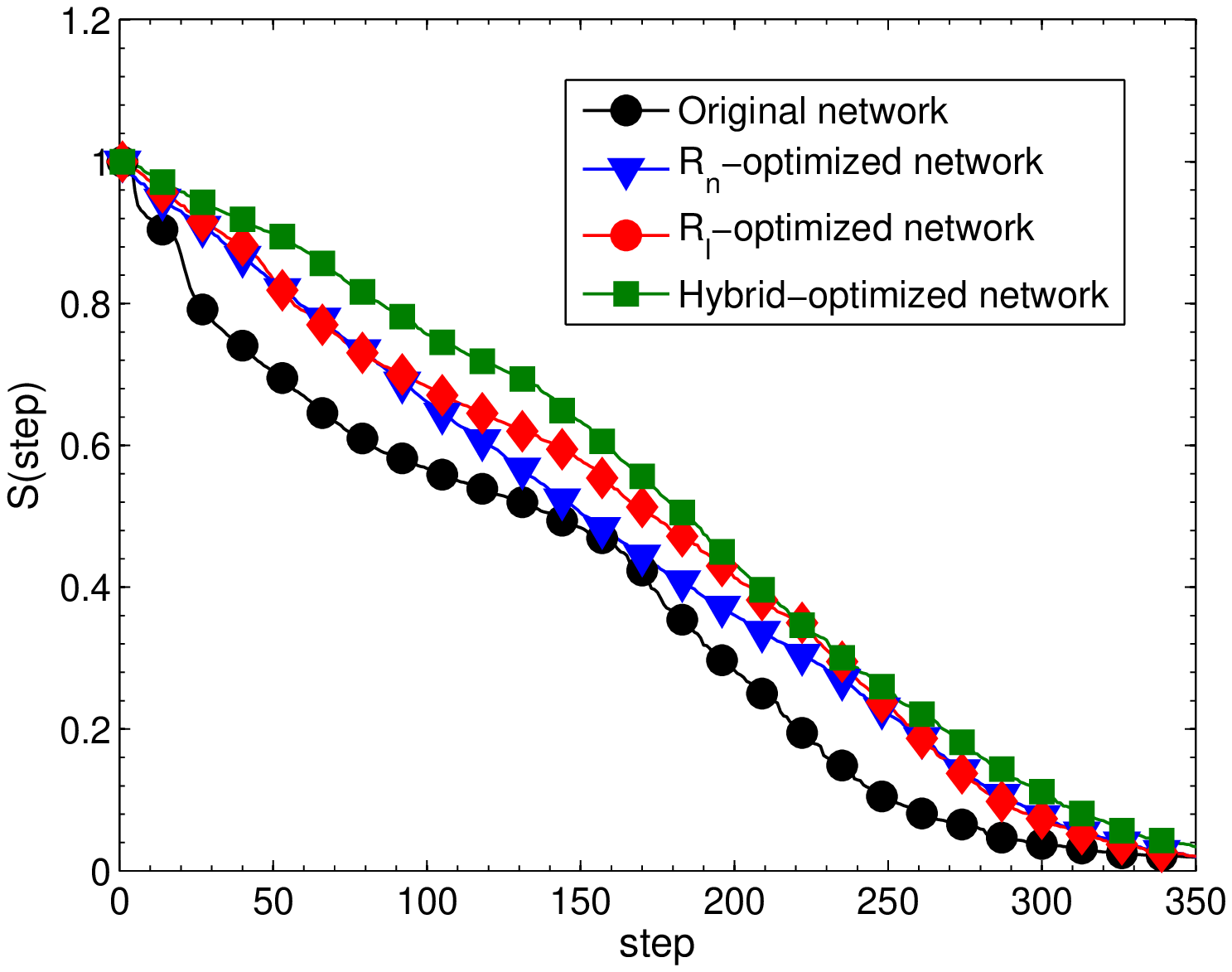}
\caption{(Color online) The change of the relative size of giant components $S$ when different networks are attacked by the mixed strategy. The original network is USAir and the fraction of node failure $f$ is set as $0.5$. The results are averaged over $100$ independent realizations. }
\label{fig3}
\end{figure}

\begin{figure}
  \center
  \includegraphics[width=7cm]{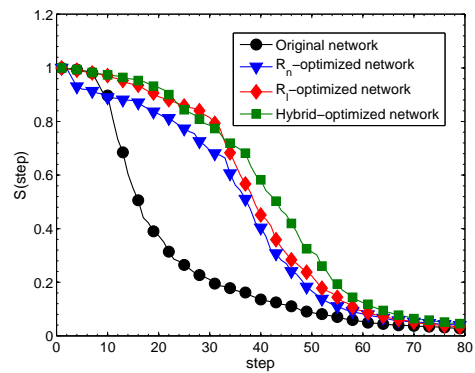}
\caption{(Color online) The change of the relative size of giant components $S$ when different networks are attacked by the mixed strategy. The original network is Grid and the fraction of node failure $f$ is set as $0.5$. The results are averaged over $100$ independent realizations.}
\label{fig4}
\end{figure}

In real systems, the failures can actually happen in not only nodes but also links. For example, heavy snow can break some power cables and aircraft mechanical problem can block certain airlines. Therefore, when designing the robust networks, we should take both $R_{n}$ and $R_{l}$ into account. In order to achieve this objective, we propose a hybrid greedy algorithm to manipulate the network structure for better robustness. Different from the process in the previous section, we swap the connections of two randomly chosen edges only if both $R_{n}$ and $R_{l}$ are improved. The swapping process stops if there is no improvement in $10^4$ trials.

Besides the BA network model, we further consider two real systems: (1) USAir: the network of US air transportation
system~\cite{USAirdata}, which contains $332$ airports and $2126$ airlines. (2) Grid: an electrical power grid of western Europe (mainly Portugal and Spain)~\cite{PowerGriddata}, with nodes representing generators, and links corresponding to the high-voltage transmission lines
between them. This network contains $217$ nodes and $320$ links. Both real networks are well connected and without any isolated component.

For each network mentioned above, we obtained the corresponding $R_{n}$-optimized, $R_{l}$-optimized and Hybrid-optimized networks by the greedy algorithm and the related results are given in table I. As we can see from the BA model and USAir network, optimizing $R_{n}$ cannot guarantee the improvement of $R_{l}$ and optimizing $R_{l}$ cannot always increase $R_{n}$ either. However, the hybrid method can improve both $R_{n}$ and $R_{l}$ from the original networks. More specifically, the $R_{n}$ and the $R_{l}$ are increased respectively by $78.2\%$ and $22.1\%$ in the USAir network. In the Grid network, the improvement of $R_{n}$ is $46.4\%$ and the increment of $R_{l}$ can reach even $154.8\%$. Compared with $R_{n}$-optimized and $R_{l}$-optimized networks, the hybrid-optimized networks do not have the advantage in single aspect of robustness, but they are kept with a reasonable balance between $R_{n}$ and $R_{l}$.

As we mentioned in the introduction, the network robustness was formerly characterized by the spectrum of the adjacency matrix ($\lambda_1/\lambda_2$), we however show that the spectrum index has certain positive correlation with $R_{n}$ but doesn't have obvious relation to $R_{l}$. Therefore, it actually only represents the node-robustness but cannot reflect the network robustness for link attack. The topology properties of the resultant networks are also analyzed. The result in table I shows that the hybrid-optimized networks usually have larger assortativity, smaller average shortest path length and lower cluster coefficient than the original networks. It has been revealed that the optimal structure for $R_{n}$ is the onion structure in which nodes with almost the same degree are connected, so the most significant feature for $R_{n}$-optimized network is the large assortativity. For the aspect of $R_{l}$, since the most destructive attack strategy is based on the highest load (edge-betweenness), the less significant the community structure is, the higher $R_{l}$ will be. Consequently, the robust network against to the link attack should be with small average shortest path length and cluster coefficient. Unlike the onion network, the $R_{l}$-optimized networks usually don't have a large assortativity, which explains why the onion network don't have a high $R_{l}$. For the resultant network from the hybrid algorithm, they will finally carry these topology properties from both $R_{n}$-optimized and $R_{l}$-optimized networks.

\begin{figure}
  \center
  \includegraphics[width=7cm]{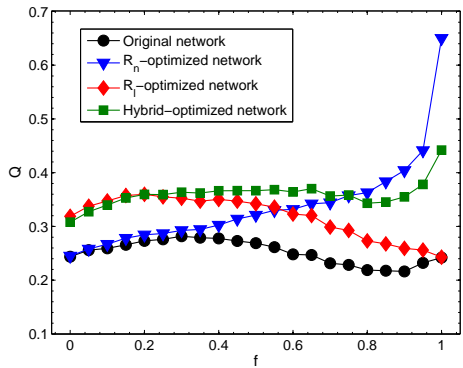}
\caption{(Color online) The $Q$ value of different networks when $f$ changes from $0$ to $1$. The original network is USAir. The results are averaged over $100$ independent realizations.}
\label{fig5}
\end{figure}

\begin{figure}
  \center
  \includegraphics[width=7cm]{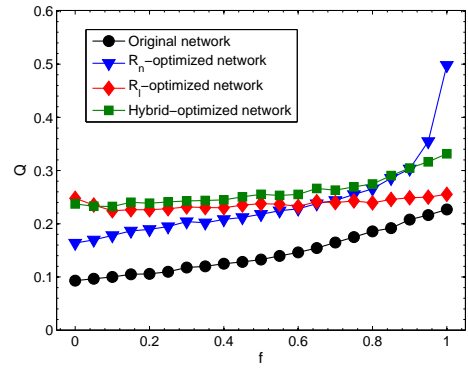}
\caption{(Color online) The $Q$ value of different networks when $f$ changes from $0$ to $1$. The original network is Grid. The results are averaged over $100$ independent realizations.}
\label{fig6}
\end{figure}

Since the failures in nodes and links can happen simultaneously, a robust network should be able to resist the attack from both ways. One interesting aspect to consider is to see how the networks in table I react to the attack combining node failures and link failures. Accordingly, we design a mixed attack strategy in which the nodes will be removed with probability $f$ and the links will be cut with probability $1-f$ in each step. The procedure goes on until the size of the giant component reaches $0$. We first set $f=0.5$ as an example and report in fig. 3 and 4 the performance of the networks in table I. The results show that the hybrid-optimized networks preserve the giant component most effectively.

We then consider the mixed attack process with $f$ varying from $0$ to $1$. When $f=0$, the process is just pure highest load (edge-betweenness) attack on links. When $f=1$, it returns to the largest degree attack on nodes. Here, we are mainly interested in the situation where $0<f<1$. In order to estimate in which range of $f$ the hybrid-optimized network has advantage, we generalize the definition of robustness to a quantity $Q$ in the mixed attack process,
\begin{equation}
Q=\frac{1}{M}\sum^{M}_{step=1}S(step),
\end{equation}
where $M$ is the total number of steps to entirely destroy the network connectivity. $Q$ measures how tolerant a network against the malicious attack (which can be nodes attack, link attack or mixed). According to Eq. (2), $Q=R_{l}$ when the $f=0$ and $Q=R_{n}$ when $f=1$.

The $Q$ value of the networks in table I under different $f$ are reported in fig. 5 and 6. Obviously, the original networks performs worst under any $f$. The $R_{n}$-optimized networks can indeed improve the $Q$ value when $f$ is large. However, they don't have too much advantage when $f$ is small. More specifically, in the USAir network (see fig. 5), the $R_{n}$-optimized network has almost the same $Q$ when $f$ is smaller than $0.4$. The $R_{l}$-optimized network can significant improve the $Q$ value when $f$ is small, but $Q$ drops nearly back to the original network level when $f$ is large. The similar trend can be seen also in the Grid network (fig. 6). These phenomena indicate that the $R_{n}$-optimized network is very sensitive to link attack while the $R_{l}$-optimized network is fragile when attacked by nodes. The hybrid-optimized networks, however, perform very stable under different attack situations (i.e., different $f$), which suggests that the hybrid-optimized network is a much more reliable structure in reality, especially when the fraction of node and link failure is unknown. Moreover, compared to the $R_{n}$-optimized and $R_{l}$-optimized networks, the hybrid-optimized network can even enjoy a higher $Q$ value in certain range of $f$ ($0.2\leq f\leq0.75$ in the USAir network and $0.1\leq f\leq0.9$ in the Grid network). In other words, when the network is attacked by both links and nodes, the hybrid-optimized network seems to be the most robust structure.

Finally, we consider some economical constraint on improving the robustness in the real system. First of all, the total length (geographically
calculated) of links cannot be exceedingly large. Secondly, the number of changes of links should be relatively small. Therefore, for reconstructing the real networks like USAir and Grid, we add two more constraints to the greedy algorithm: the swap
of two links is only accepted if the total geographic length of edges does not increase, and both $R_{n}$ and $R_{l}$ are increased more than certain values (denoted as $\Delta R_{n}$ and $\Delta R_{l}$)~\cite{explanation}. With the strong constraints, $R_{n}$ and $R_{l}$ of real networks can still be significantly improved. Specifically, with only $3.9\%$ links changed, the $R_{n}$ and $R_{l}$ of the USAir network can be respectively increased by $56\%$ and $17\%$ ($R_{n}$: from $0.110$ to $0.172$. $R_{l}$: from $0.244$ to $0.285$). In the Grid network, the $R_{n}$ can be improved by $23\%$ (from $0.111$ to $0.136$) and the $R_{l}$ can be improved by $20\%$ (from $0.093$ to $0.112$) with only $6.9\%$ links changed.

\section{Conclusion}

How to enhance the robustness of networks is an important topic, which is related to protecting the real system from random failures and malicious attacks. In the former literatures, most of the works focused on proposing methods to improve the network robustness for the attack on nodes. However, the connections between nodes can be also damaged due to some unexpected accidents, which requires us to take the link failure into account when designing robust networks. In this paper, based on the highest load attack strategy, we propose the link-robustness index to estimate how the network can resist to the most destructive targeted attack on links. Moreover, we designed a hybrid greedy algorithm to enhance both node-robustness and link-robustness. When attacked by the strategy combining node and link failure, the resultant networks from the hybrid method outperform the networks from solely improving either $R_{n}$ or $R_{l}$. Finally, some economical constraints are considered when enhancing the robustness of real networks and some significant improvement are observed.

Although the hybrid method can obtain a reliable network which is generally robust to the attack mixed with node failures and link failures, there are still some further improvement can be achieved. In real system, the probability of the node failure and link failure can be different from one system to another. In the paper, we accept the swap of links only if both $R_{n}$ and $R_{l}$ are increased. However, one can sacrifice $R_{n}$ a little bit for larger improvement in $R_{l}$ or the other way around since these two robustness aspects ask for different structure properties. Provided knowing the fraction of node failure and link failure from the analysis of the historical data, more effective greedy algorithm can be designed to generate suitable network structures for some specific real systems.

\section*{acknowledgement}
We would like to thank Yi-Cheng Zhang and Xiao-Pu Han for helpful suggestions. This work is supported by the Swiss National Science Foundation
(No. 200020-132253).

\end{document}